\documentclass[aps,prb,reprint,groupedaddress]{revtex4-1}

\usepackage{graphics}
\usepackage{epsfig}

\begin{document}

\preprint{}

\title{Experimental determination of correlations between spontaneously formed vortices in a superconductor}


\author{Daniel Golubchik}
\email[danielg@tx.technion.ac.il]{}
\affiliation{Department of Physics, Technion - Israel Institute of Technology,
Haifa 32000, Israel}
\author{Emil Polturak}
\affiliation{Department of Physics, Technion - Israel Institute of Technology,
Haifa 32000, Israel}
\author{Gad Koren}
\affiliation{Department of Physics, Technion - Israel Institute of Technology,
Haifa 32000, Israel}

\author{Boris Ya. Shapiro}
\affiliation{Department of Physics, Bar Ilan University,
Ramat Gan, 52900, Israel}

\author{Irina Shapiro}
\affiliation{Department of Physics, Bar Ilan University,
Ramat Gan, 52900, Israel}


\date{\today}

\begin{abstract}
We have imaged spontaneously created arrays of vortices (magnetic
flux quanta), generated in a superconducting film quenched through
its transition temperature at rates around $10^9 K/s$. From these
images, we calculated the positional correlation functions for two
vortices and for 3 vortices. We compared our results with
simulations of the time dependent Ginzburg Landau equation in 2D.
The results are in agreement with the Kibble-Zurek scenario of
spontaneous vortex creation. In addition, the correlation
functions are insensitive to the presence of a gauge field.
\end{abstract}


\maketitle

Vortices (magnetic flux quanta) are topological defects of the
order parameter in a superconductor. These defects are predicted
to appear spontaneously, as a result of a rapid quench through the
phase transition into the superconducting state. During a rapid
quench, close to the transition temperature ($T_c$), the
relaxation time of the system becomes larger then the quench time.
Under these conditions, the system is necessarily driven out of
equilibrium. One model that describes the outcome of such phase
transition is the Kibble-Zurek scenario, first suggested by
Kibble\cite{Kibble:1976,Davis:2005} in a cosmological context.
Among his other contributions, Zurek \cite{Zurek:1985} proposed
terrestrial tests of this model in condensed matter systems, where
the broken symmetry is U(1), such as superfluids, BEC and
superconductors. The behavior of these systems can be described by
a complex order parameter. Above the critical temperature $T_c$,
the order parameter fluctuates with a characteristic size of the
fluctuation being $\xi$. Is the sample is cooled infinitely slowly
towards $T_c$,  $\xi$ will grow until it reaches the size of the
sample. At finite cooling rates the fluctuations "freeze" at some
point, forming isolated, uncorrelated domains of the ordered
state. The typical size of such a domain, $\hat{\xi}$ , inside
which the emerging order parameter is coherent, depends on the
cooling rate. In equilibrium, the order parameter in the final
state should be uniform across the system. However, the initial
mismatch of the phase of the order parameter  between different
regions leads to the appearance of topological defects. In
superconductors, these are vortices carrying a quantum of magnetic
flux $\Phi_0 \equiv h/2e$.

In the KZ model, below $T_c$ uncorrelated domains of the ordered
state are formed. Each of these domains picks up a random value of
phase of the order parameter. Single-valuedness of the order
parameter requires that the integral of the phase accumulated
along the circumference of any loop must be an integer multiple of
$2\pi$. If the phases of different domains are random, there is a
finite probability that the phase accumulated along a loop around
a vertex between 3 domains will be $\pm2\pi$, leading to formation
of topological defect.  To calculate the probability of this
formation, the geodesic rule is usually implemented. It assumes
that minimal phase gradient will always be favorable due to
minimal energy consideration. Under this assumption, only one
vortex can be created at a vertex between 3 ordered regions.
Hence, the geodesic rule restricts the number of topological
defects created by the system. This description is limited to
relaxation of the phase gradients and does not involve any
additional dynamics. One of the predictions of this model is
strong, short range correlation between vortices and
anti-vortices. If the size of frozen fluctuations $\hat{\xi}$ is
assumed to have a gaussian distribution, one can calculate the
vortex-vortex correlation function \cite{Liu:1992}. Hence, the
fluctuations distribution above the critical temperature leaves
its mark on the emerging vortex array. By measuring the positions
of the vortices and calculating the correlation function it is
possible to investigate the fluctuation distribution above $T_c$.

We note that the KZ model does not address the critical coarsening
process which may occur after the transition \cite{Biroli:2010}.
In many systems, coarsening affects drastically the outcome of the
transition due to defect-antidefect annihilation. This may not be
the case for superconducting films. If the quench is fast enough,
the system reaches low temperatures before vortices can traverse
the distance to a nearby antivortex and annihilate. At
temperatures much lower than $T_c$ vortices are strongly pinned,
so that any motion is practically impossible. As a result, vortex
annihilation and coarsening is suppressed. Superconductors
therefore offer us an unique opportunity to investigate the order
parameter fluctuations above $T_c$, by measuring the vortex
distribution after a quench.

The validity of KZ mechanism in systems with local gauge
symmetries (such as superconductors), has been questioned by
several authors \cite{Rudaz:1993}. For these systems an
alternative mechanism of flux trapping was suggested
\cite{Hindmarsh:2000,Kibble:2003,Donaire:2007}. In this mechanism
thermal fluctuations of the magnetic field are frozen inside the
superconductor during the transition.  As a consequence vortices
are formed in clusters of equal sign. The resulting vortex-vortex
correlation function should decay as a power
law\cite{Rajantie:2009}. However, the amplitude of trapped
magnetic field fluctuations in conventional superconducting films
should be so low that vortex formation should be better described
by the Kibble–Zurek mechanism\cite{Donaire:2007}.

It was suggested by several authors that topological defects in
first order phase transitions
\cite{Copeland:1996,Digal:1996a,Digal:1996b,Donaire:2009} and in
sustems showing spinodal decomposition \cite{Pezzutti:2009} may
form due to dynamics. In this approach, the geodesic rule does not
hold anymore. Pairs of vortices and anti-vortices are formed
during collisions between domain walls. Although most of the
simulations were done for first order phase transitions, this
mechanism is claimed to be generally applicable \cite{Digal:1997}.
The correlation function for this mechanism was never calculated,
but it should have two characteristic lengths: the separation
between nearby vortices produced by domain wall collisions, and
the domain size.

Vortex pairs of opposite sign could also arise from a
Kosterlitz-Thouless type of transition\cite{Chu:2001}. In this
theory, unbound vortex pairs appear above $T_{KT}$. If the system
is cooled through $T_{KT}$, these vortices effectively annihilate.
However, if the quench is fast, some of the unbound vortex pairs
can survive the quench, become pinned at low temperatures and
observed. In this theory, the density of unbound vortex pairs
above $T_{KT}$ increases. Consequently, the observed vortex
density should depend on the temperature from which the system is
quenched. Within our resolution, we found no such dependence in
the experiment described here.

The Kibble-Zurek (KZ) model has been tested in liquid helium
\cite{Bauerle:1996,Ruutu:1996}, liquid crystals
\cite{Rajarshi:2004}, in superconductors \cite{Maniv:2003},
Josephson junctions \cite{Carmi:2000,Monaco:2002} and
superconducting loops\cite{Kirtley:2003,Monaco:2009}.
Spontaneously generated topological defects were detected in
several of the aforementioned experiments. More sensitive testing
of the model involves the determination of correlations between
these defects. There are only two such experiments performed to
date. In one experiment\cite{Rajarshi:2004}, done with liquid
crystals, an array of topological defects was imaged. However, the
amount of data was insufficient to detect correlations beyond
nearest neighbors. In our experiment\cite{Golubchik:2010}, we
imaged spontaneously formed arrays of vortices in a
superconductor. The amount of data gathered was large, allowing a
precise determination of the two point correlations. In this work,
we extend this study to look at correlations between 3 defects.

Our experimental technique is described in detail in previous
publications\cite{Golubchik:2009,Golubchik:2010}. Briefly, we
image the magnetic field on the surface of a superconducting film
using high resolution Magneto-optics. The superconductor sample
consists of a $200 nm$ thick Niobium film with $T_c$ of $8.9K$.
The film is patterned into small squares of $200\mu m$ across. On
top of the Nb film, we deposited a $40 nm$ layer of $EuSe$ which
serves as the Magneto-Optic sensor. To minimize the effect of
stray magnetic fields, our apparatus was carefully shielded using
$\mu$-metal. The residual field was less than $10^{-7} T$).

From our previous experiments \cite{Maniv:2003}, we know that
extremely high cooling rates are essential for spontaneous
generation of a measurable amount of vortices. No less important,
fast cooling to low temperatures (far below $T_c$) traps the
vortices on pinning centers, preventing annihilation of vortices
and anti-vortices. Using short laser pulses\cite{Golubchik:2010},
we achieved cooling rates as high as $2\cdot10^9 K/s$.

\begin{figure}
\includegraphics[width=3.2093in]{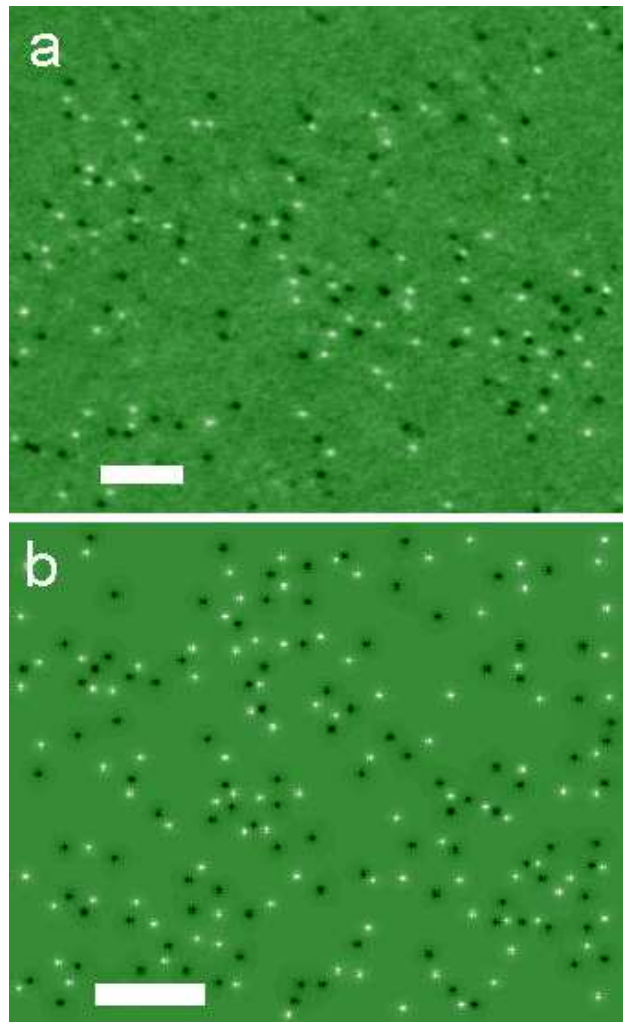}
\caption{Spontaneously created vortices in a
superconductor. \textbf{a}) A typical image of magnetic field
created after a quench. The intensity is proportional to the local
magnetic field. Bright and dark spots are vortices and
anti-vortices respectively. The scale bar represents 10 $\mu m$.
\textbf{b}) A typical result of the simulation of TDGL equations.
The scale bar represents 10 $\xi_0$. \label{spont}}
\end{figure}

Fig.1a shows a typical image of spontaneously generated vortices.
The average asymmetry between the density of positive and negative
vortices is less than $1\%$. The  average density of vortices was
$6\cdot10^5 cm^{-2}$ for a cooling rate of $4\cdot10^8 K/s$ and
$1.3\cdot10^6 cm^{-2}$ for a cooling rate of $2\cdot10^9 K/s$. The
scaling of the density with the cooling rate is consistent with
the KZ model at 2D (proportional to the square root of the cooling
rate).

\ Our experimental results are compared with numerical simulations
of the 2D time dependent Ginzburg-Landau equations. The
simulations are described in detail in \cite{Ghinovker:2001}.
Briefly, we used a square $200\times 200$ grid, with the
parameters $\kappa =1$, $\Gamma =1$. The initial conditions, $A=0,
\Psi=0, \theta =0.7$, mimic a quench from temperatures far above
$T_c$ to $T=0.7 T_c$.   The implicit Crank-Nicholson scheme was
employed on a staggered grid with step in time $\Delta \tau =0.01$
and in space $\Delta_l=0.2$. The time of each run was $20\tau
_{GL}$. After this time the vortices are well defined and assumed
to be pinned. A typical result is presented in fig.1b. Bright and
dark spots mark the positions of vortices and anti-vortices.
Several of the parameters of the simulation are different from the
experimental parameters. First, the cooling rate in the simulation
is infinite, in contrast to the finite rate in the experiment.
Finite cooling rates in the simulation scale the correlation
length but do not affect the distribution. Second, the simulation
is fully two dimensional, while the sample used in the experiment
is a thin film. The interaction between vortices in thin films is
much stronger, and mediated through the magnetic field of the
vortex outside of the film \cite{Brandt:2009}.

\begin{figure}
\includegraphics[width=3.2093in]{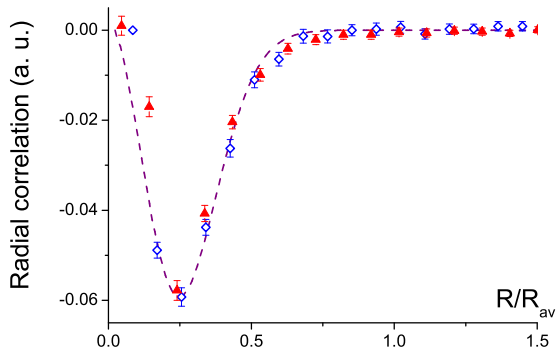}
\caption{The vortex-vortex correlation function
$G(r)$. Solid red triangles represent the correlation
function calculated from experimental data. The statistical error
bars are smaller than the point size. Open blue diamonds
represent correlations calculated from the results of simulation.
The solid line is a fit to theory of Liu and
Mazenko\cite{Liu:1992}. The negative peak at short distance
reflects vortex-antivortex correlations predicted by KZ model.
\label{vortcor}}
\end{figure}

We used images like those shown in Fig.1 to determine the
correlations between the vortices. The two particle vortex-vortex
correlation function is defined as
$G(r-r^\prime)=<n(r)n(r^\prime)>$ , with $n(r)=1$ at the location
of a positive vortex, $-1$ at the location of negative vortex and
$0$ elsewhere. The correlation function calculated from our data
is shown in Fig.2. The distance in the figure was scaled by the
mean vortex separation, $r_{av}=<\rho>^{-1/2}$ as proposed by
\cite{Rajarshi:2004}. $r_{av}$ is related to $\hat{\xi}$ by
$r_{av}=\frac{\hat{\xi}}{\sqrt{p}}$, where $p$ is the average
number of vortices per domain. For our high cooling rate of
$2\cdot10^9 K/s$, $r_{av}$ is $8.2 \mu m$. The correlation
function was averaged over 260 images, with 50,000 vortices in
total. At the distance corresponding to the nearest neighbors the
correlation function has a minimum. This is consistent with the KZ
model which predicts that nearest neighbor vortices should have
opposite polarities. The correlation function calculated from the
simulations in the Fig.2 was averaged over 200 realizations. Since
the same scaling was used, the experimental and simulated
correlation functions can be compared without additional
parameters. The dashed line is a fit to the theoretical
predictions by Liu and Mazenko\cite{Liu:1992} with $\hat{\xi}=0.35
r_{av}$. In their work, Liu and Mazenko assumed a Gaussian
distribution of the fluctuations. Even through the simulation
parameters differ from our experimental conditions (infinite vs.
finite cooling rate, different initial temperatures, different
interactions between vortices), the resulting correlation function
is the same. Qualitatively similar correlation function was also
calculated from simulations of quenched 2D XY
model\cite{Jelic:2010}. The agreement with the correlation
function calculated by Liu and Mazenko\cite{Liu:1992} suggests
that in all cases, the shape of the correlation function is
dictated by Gaussian fluctuations.

\begin{figure}
\includegraphics[width=3.2093in]{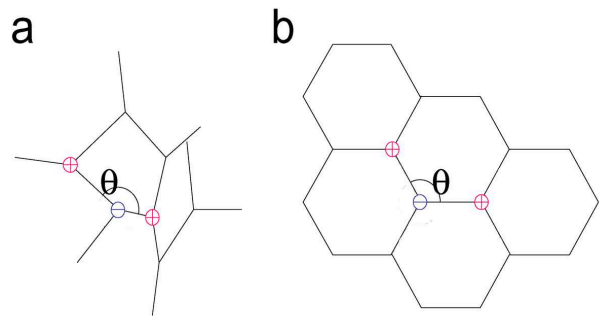}
\caption{Schematic description of spontaneous
formation of vortices on intersections between ordered domains. a)
Random domain array. b) Close packed domain structure. Vortices
and anti-vortices are marked by + and - respectively.
\label{ang_dom}}
\end{figure}

In addition to the two point correlation function $G(r-r^\prime)$,
we used our data to calculate an angular ( 3 point) correlation
function. This function involves 3 vortices. For each vortex in
the array, we look for its two nearest neighbors. Nearest
neighbors are defined as vortices located within a distance of
$\sim \hat{\xi}$ and have an opposite polarity (see fig.3). If
such neighbors are found, we calculate the angle between the two
lines connecting the vortex to its neighbors. In the limit where
the domains are mono-dispersed and close packed (fig.3b), the angle
between nearest neighbors will be $120^o$. For randomly
distributed domains (fig.3a) the angles will have a broad
distribution. Even through the connection between the fluctuation
distribution and the angle distribution was never calculated, the
first obviously determines the second. Therefore the comparison
between the angle distribution calculated out of experimental
data, and from the results of a simulation can be used as a
validity test for a model. The angle distributions calculated from
experimental data and from simulation are shown in Fig. 4. Both
distributions have a minimum at $\theta=0^o$ and increase until
$\theta\approx100^o$ were they reach a plateau. Within the error
bars, these two distributions are consistent.

\begin{figure}
\includegraphics[width=3.2093in]{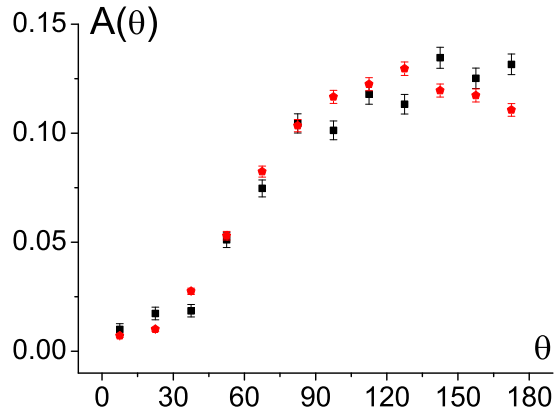}
\caption{ The angular distribution function. Solid black
squares represent the angular distribution
calculated from experimental data. Red pentagons
are angular distribution calculated from the results of the
simulation. \label{angcor}}
\end{figure}

In conclusion, we calculated the vortex-vortex correlation
function and the angular distribution function using experimental
data and compared them with the results of TDGL simulations. We
found an agreement between the experimental and simulated results.
This implies that within experimental accuracy, the TDGL equation
in 2D describes the vortex formation in our system. One
consequence is that the coupling to the gauge field outside the
sample does not affect the vortex distribution, at least for the
parameters used in the experiment. The correlation function
calculated using a quenched XY model\cite{Jelic:2010} gave
qualitatively similar results, which suggests that gauge fields
have no effect on the system evolution during quench.


\begin{acknowledgments}
We thank S. Lipson and E. Buks for their contribution to this
experiment. We thank S. Hoida, L. Iomin and O. Shtempluk for
technical assistance. This work was supported in part by the
Israel Science Foundation (Grant  No. 499/07) and by the Minerva and DIP projects.
\end{acknowledgments}



\end{document}